\documentstyle[aps]{revtex}
\begin{document}
\def\beq{\begin{eqnarray}}
\def\eeq{\end{eqnarray}}

\title{Skirting Hidden-Variable No-Go Theorems}
\draft
\author{A. F. Kracklauer}
\address{Belvederer Allee 23c \\ 99425 Weimar, Germany 
\\ kracklau@fossi.uni-weimar.de}
\date{\today}

\twocolumn[           
\maketitle
\begin{minipage}{\textwidth}   
\begin{quotation}           

\begin{abstract}
It is observed that the proofs of hidden-variable no-go theorems 
depend on the `projection postulate,' which is seen to be 
contradictory with respect to spin operators in directions orthogonal 
to the magnetic field direction.  In this light it is argued  that it 
is less costly to abandon the projection postulate than to abandon 
locality and this in turn renders Hidden-Variable No-Go Theorems 
evadible.  To buttress this point, a local realist model of the EPR 
experiment which ignores the constraints of the projection postulate 
is presented. 

\bigskip   
\end{abstract} 
\end{quotation}      
\end{minipage}]      

\section{INTRODUCTION}

In the Theory of Quantum Mechanics (QM), as is well known, 
complementary variables can not be determined to arbitrary precision. 
This fact in combination with certain symmetries, however, leads 
directly to a contradiction.  Consider, for example, the 
disintegration of a stationary particle into twin daughters moving off 
in opposite directions.  For each daughter separately, only the 
position or momentum can be measured with arbitrary precision.  Now 
however, if at a given instant the position of one daughter and the 
momentum of the other are determined precisely, then by symmetry, the 
complementary position and momentum for both twins can be deduced  
with arbitrary precision, seemingly in conflict with QM.  These 
considerations clearly imply that in spite of a QM dictum, 
complementary variables must have a certain `reality' independent of 
that which should be created by a measurement. 

Indeed, in 1935 Einstein, Podolsky and Rosen (EPR) used exactly this 
argument to challenge the assumed completeness of QM.  EPR 
anticipated the possibility that the insinuation of additional, 
heretofore `hidden,' variables could render QM complete; that is, with 
the aid of these extra variables the calculation of even complementary 
quantities could be done, at least in principle, with arbitrary 
precision.  In fact, inspired by the above arguments, EPR used the 
characteristic of being predictable with certainty as the very 
definition of `reality.' \cite{EPR}  Accordingly, attempts to discover 
these extra, still hypothetical, hidden variables have been identified 
with a so called `realist' school. 

Since that time Von Neumann \cite{Neumann}, Bohm \cite{Bohm}, Bell 
\cite{Bell1} and many many other researchers have with various 
tacks studied the issue. \cite{Mermin} In spite of, perhaps because 
of, the essential ambiguity of the interpretation of QM, some sought 
refuge in the rigor of mathematics.  But, within the discipline of 
Physics it is impossible to prove a theorem.  The proof of a theorem 
requires a set of established theorems and definitions built up from a 
base of axioms.  This is possible in Mathematics and Logic but the 
Natural Sciences are different.  They are programs to find the  
`axioms of Nature,' as it were.  These axioms certainly are largely 
unknown (else basic research could be abolished in favor of 
engineering, etc.), and candidate axioms, in the form of fundamental 
theories, currently under discussion are not mutually  fully 
consistent.  Lacking axioms, there can be no web of syllogisms, hence 
no proven theorems, just theories with their intrinsically tentative 
nature. 

In this light one may ask just how has Bell's `Theorem,'---to the effect 
that efforts by `realists' to identify `hidden' variable are doomed to 
finding only such which are nonlocal; i.e., those entailing 
dependency on events outside past light cones---been proven?  
Rummaging in the literature eventually discloses that for this purpose 
an implicit axiom set has been mandated.   As this set comprises the 
formal structure of QM and not just empirical data or the phenomena to 
be described, the physical justification for these axioms is open to 
suspicion.  In particular, and this will turn out to be the nub of the 
argument made herein, it includes the `projection postulate' or the 
assertion that an individual measurement on a single system in effect, 
always and exclusively, `projects' out the eigenvalue of an operator 
on an appropriate Hilbert space as a result.  Without this postulate, 
or at least parts of its operational equivalent, proofs of Bell's 
Theorems do not go through. It will be argued below that imposing this 
postulate on hidden variable formulations of QM restricts such 
theories in a way in which the core of QM itself in fact is not. The 
cost of abandoning this postulate, however, is trivial compared with 
that of abandoning locality. 

As regards a `hidden variable' or `realist' program, such an 
adaptation is in harmony with the motivation of its proponents.  Their 
goal was and is to explain `Nature' not QM.  If this can be done 
successfully without the full generality of the projection postulate, 
but otherwise in accord with the empirically verified results of QM, 
there can be no objection. 

That the possibility of evading Bell's results exists was telegraphed 
already in 1980 by the late Pierre Claverie who found a mathematical 
gremlin in the proof.  \cite{Claverie} His point is that dichotomic 
functions representing spin cannot be correlated to yield a harmonic 
function, which is the so called `QM-result' (Eq. (14) below).  
The question of just how general this objection is, and whether it 
extends  to all variations of Bell's Theorem remains open.  It is the 
point of this note to address this issue. 

\section{SOME FUNDAMENTALS}

We recall here some elementary facts concerning spin.  Although well 
known, they are essential below and for convenience are 
repeated.\cite{Dicke}   

Given a uniform static magnetic field ${\bf B}$ in the $z$-direction, 
the Hamiltonian is:

\beq
{\bf H}=+{e\over{mc}}{\bf B}{\sigma_z}. 
\eeq

For this Hamiltonian, the time-dependent solution for the 
Schr\"odinger equation is: 

\beq
\psi(t)={1\over{\sqrt{2}}}\left[\matrix{ e^{-i\omega 
t\over{2}} \cr e^{+i\omega t\over{2}} } \right], 
\eeq

and it gives time-dependent expectation values for spin values in the 
$x,y$ directions respectively: 
\beq\langle\sigma_x\rangle = {\hbar\over{2}}\cos(\omega 
t),\,\,\langle\sigma_y\rangle = {\hbar\over{2}}\sin(\omega t),\eeq 
where $\omega = e{\bf B}/{mc}$.

What is to be seen here is that in the $x$-$y$ plane, in contrast to 
the $z$, magnetic field, direction, expectation values can not be 
made up out of averaged eigenvalues.  In this plane a direct 
application of the projection postulate seems absurd.  In fact, what 
is usually meant by measuring spin in these directions requires 
reorienting the magnetic field so that in effect one again is 
measuring $\sigma_z$ in the new $B$-direction.

\section{HIDDEN-VARIABLE NO-GO THEOREMS}

There are two basic formulations of Bell's hidden-variable no-go 
theorems: a version found first by Bell \cite{Bell2} but shortly 
thereafter refined and generalized by Kochen and Specker \cite{Kochen} 
and a second version found by Bell \cite{Bell3} and subsequently 
modified by Clauser to accommodate experimental exigencies. 
\cite{Clauser2} 
                     
The latter version is the one most often discussed in the literature. 

\subsection{Bell-Kochen-Specker Theorems}

For the purposes of this note, I call on a version of the 
Bell-Kochen-Specker Theorem expounded by Mermin.  \cite{Mermin} At the 
onset, the system of interest is presumed prepared in a `state' 
$|\psi\rangle$ and described by observables $A,B,C\dots$. A hidden 
variable theory is then deemed to be a mapping $v$ of the observables 
to numerical values:  $v(A), v(B), v(C) \dots$ so that if any 
observable or mutually commuting subset of observables is measured on 
that system, the results of the measurements on it will be the 
appropriate values.  Just how the values are fixed is the substance of 
the particular theory. 

Now if this theory is to be compatible with QM, the observables are 
operators on a Hilbert space (An erudite way to say that the solutions 
of the Schr\"ordinger Equation reflect its hyperbolic nature.) and 
that the measured values $v(A)$, etc. are the eigenvalues of these 
operators.  It is simply a fact that if a set of  operators all 
commute, then any function of these operators $f(A,B,C \dots)=0$ will 
also be satisfied by their eigenvalues: $f(v(A), v(B), v(C) \dots)=0$. 

From this point, the proof of a Kochen-Specker Theorem proceeds by 
displaying a contradiction.  Surely the least complicated rendition 
of the proof considers two `spin-$1/2$' particles. For these two 
particles, nine separate mutually commuting operators are formed into 
the following 3 by 3 matrix:
\beq
\begin{array}{ccc}
\sigma_x^1 & \sigma_x^2 & \sigma_x^1 \sigma_x^2 \\[2mm]
\sigma_y^2 & \sigma_y^1 & \sigma_y^1 \sigma_y^2 \\[2mm]
\sigma_x^1 \sigma_y^2 & \sigma_x^2 \sigma_y^1 &
          \sigma_z^1 \sigma_z^2.
\end{array}\eeq

It is a little exercise in bookkeeping to verify that any assignment 
of plus and minus ones for each of the factors in each element of this 
matrix results in a contradiction, namely, the product of all these 
operators formed row-wise is plus one and the same product formed 
column-wise is minus one.

Here we see immediately that the proof depends on simultaneously 
assigning the [eigne]values $\pm 1$ to $\sigma_x$, $\sigma_y$ and 
$\sigma_z$ as measurables for each particle.  (With little effort, for 
all other proofs of this theorem one can find the same assumption.)  
However, in Section II above, we saw that if the eigenvalues $\pm 1$ 
are believable measurable results in one direction, then in the other 
two directions the expectation values oscillate.  Thus, in the 
oscillating directions, the projection postulate makes no sense.  (It 
appears as if it should at least be limited to operators for which the 
matrix of eigenvectors is diagonalized.  Moreover, it has been 
suggested that the projection postulate is generally invalid. 
\cite{Ferrero} This author has great sympathy for this assertion on 
the grounds that ontologically realizable states should yield positive 
definite Wigner densities. \cite{Paradigm} Such considerations are 
beyond the limited scope of this note, however.) If the projection 
postulate is abandoned for these directions, then the proof of a Bell-
Kochen-Specker theorem does not go through.  Sacrificing it for $x$-
$y$ spin operators has no practical cost for QM---in fact, it seems 
necessary for internal consistency.  Moreover,  I note that this 
proof of this theorem makes no specific use of the properties of 
hidden-variables, so that it in principle pertains to QM as currently 
formulated; it speaks more to the viability of the projection 
postulate than of hidden variables. 

\subsection{Bell's Famous Theorem}

The formulation of Bell's more widely  known version of a no-go 
theorem provides constrains that are billed as unavoidable by all 
realist-local theories and which are violated by QM.  Once again, 
however, the derivation depends on implicit hypothesis including the 
projection postulate for more than one direction at a time.  To show 
this explicitly, consider Bell's seminal paper \cite{Bell3} where 
the argument proceeds as follows.  

One considers two functions $A({\bf a},\lambda)$ and $B({\bf 
b},\lambda)$ which are to represent measurements, in the case of an 
EPR experiment, the measurement of the polarization of `photons' 
emerging from a process wherein the emitted radiation can carry off no 
angular momentum; i.e., the photons must be oppositely polarized.  As 
is well known the singlet wave function for this situation is 
ambiguous with respect to the polarization.  In an effort to resolve 
this ambiguity and invest in the wave function complete predictability 
(reality), a set of heretofore `hidden' variables, $\lambda$ is 
insinuated.  This is the `reality' stipulation.  Further, as these 
measurements are to be independent of influences outside past light-
cones, $A({\bf a},\lambda)$ is to be independent of ${\bf b}$; i.e., 
variables pertaining to the measuring apparatus at the location of 
$B({\bf b},\lambda)$, which likewise is independent of ${\bf a}$.  
This is the `locality' stipulation.   

Finally, each of the functions is allowed to take the values $\pm1$ to 
correspond to the supposition that exactly two states of polarization 
are to be found.  This is the way in which the projection postulate is 
imposed on a hidden variable theory even if this theory in its final 
form should not make use of operators on a Hilbert space.  While this 
final assumption seems quite innocent, it entails many concealed 
assumptions as in fact all that can be observed is a photoelectric 
generated electron, not a distinct `photon,' for example.  As these 
functions take on the values $\pm1$ for all directions, e.g., ${\bf 
a}$ and ${\bf b}$, our point is actually complete here.  But to 
exhibit precisely how this assumption enters into the derivation of 
Bell inequalities, I continue. 

The correlation of these function then is: 
\beq
{\bf P(a,b)} =  \int d\lambda \rho(\lambda) A({\bf a},\lambda) 
B({\bf b}, \lambda), 
\eeq 
which is asked to duplicate the QM result: 
\beq
\langle {\bf \sigma^1 \cdot a\,\sigma^2 \cdot b} \rangle=-{\bf 
a}\cdot {\bf b}.  
\eeq 

To derive the famed inequalities, one then uses $A({\bf a},\lambda) 
= -B({\bf a},\lambda)$,  in the expression

\beq  
\lefteqn{P({\bf a,b})-P({\bf a,c})   = } \nonumber \\
& &  -\int d\lambda\rho(\lambda)\left[ A({\bf a},\lambda) 
A({\bf b},\lambda) - A({\bf a},\lambda)A({\bf c},\lambda)\right],  
\eeq            

which is then factored

\beq 
\lefteqn{P({\bf a,b})-P({\bf a,c}) = } \nonumber \\
& & \int d\lambda\rho(\lambda) A({\bf 
a},\lambda) A({\bf b},\lambda)\left[ A({\bf b},\lambda)A({\bf 
c},\lambda) -1 \right],   
\eeq 

where explicit use is made of 
$A({\bf b},\lambda) A({\bf b},\lambda)=1, \forall\, {\bf b}$, 
(i.e, the projection postulate) to get

\FL
\beq
\vert P({\bf a,b})-P({\bf a},c)\vert \le \int d\lambda 
\rho(\lambda) \left[1- A({\bf b}, \lambda) A({\bf c},\lambda) \right], 
\eeq 

which gives

\beq 
1+ P({\bf b,c}) \ge  \vert P({\bf a,b})-P({\bf a,c})\vert ,
\eeq

which is one variation of a Bell inequality.  Alternate derivations of 
similar inequalities employ the projection postulate in a similar way. 

\section{CORRELATIONS, Quantum vs. Local-Realist}

If in fact it is problematic to ascribe dichotomic values to spin in 
more than one direction simultaneously, the question naturally arises 
regarding just what the oft performed and empirically verified 
correlation calculation in QM textbooks accomplishes and how does it 
do it.  The answer to the above question then is that underneath the 
abstraction and notation of QM, the calculation is actually only a 
rendition of Malus' Law from classical optics applied to an ensemble 
of systems.  This is best seen by examining the derivation of the 
formulas used to compare calculations with observations. 

To begin, one considers the singlet state for a pair of `photons' 
as is usual for analysis of the Einstein-Podolsky-Rosen-Bohm 
experiment:

\beq
\psi = {1\over{\sqrt{2}}}\left[ \vert x_1 \rangle\vert y_2 \rangle-
          \vert x_2 \rangle \vert y_1 \rangle \right], 
\eeq

where here the indices $x_1, y_1$, etc. indicate photons polarized in 
the $x, y$ direction respectively and propagating away from each other 
along the $z$ direction.  The photons are then  analyzed (measured) by 
passage through, say, Wollaston prisms which separate the polarization 
states geographically.  Of this it can be said: `They perform 
dichotomic measurements, i.e., a photon can be found in one of the two 
exit channels, labeled $+1$ or $-1$. This is similar to a Stern-
Gerlach filter acting on spin $1/2$ particles.'  

For QM calculations now one transforms this state, Eq. (11), to a 
rotated frame using the standard rotation matrix:
\beq
\left[\matrix{\vert x_1'\rangle \cr \vert y_1' \rangle}\right]= 
\left[ \matrix{\cos(\phi) & -\sin(\phi) \cr                  
               \sin(\phi) & \cos(\phi)}  \right]
\left[ \matrix{\vert x_1\rangle \cr \vert y_1 \rangle} \right],
\eeq
so that for $\psi$ in the rotated frame, one gets:
\beq
\psi' & = & {1\over{\sqrt{2}}} \left[\right. \cos(\phi)\vert x_1' 
\rangle\vert y_2' \rangle  -\cos(\phi)\vert y_1' \rangle\vert 
x_2' \rangle   \nonumber \\ 
&    &  -\sin(\phi)\vert x_1' \rangle\vert x_2' \rangle -
  \sin(\phi)\vert y_1' \rangle\vert y_2' \rangle 
\left.\right]. 
\eeq 

From this expression, the essentials for QM calculations follow 
directly.  The crosscorrelation is computed as follows:
\beq
\psi'^*M_1 M_2 \psi' = -\cos(2\phi),
\eeq
where $M_1$ is a `measurement' made on photon 1; i.e., $M_1 \vert x_1 
\rangle=+1 \vert x_1 \rangle,\,\, M_1 \vert y_1\rangle=-1 \vert 
y_1\rangle$.   The probability  that both photons are polarized in the 
$x'$-direction, for example, is the square of the third term:
\beq 
P(x_1',x_2')={1\over{2}}\sin^2(\phi).
\eeq

This is Malus' Law with an extra factor of $1/2$.  Both the factor of 
$1/2$ and the apparent failure here to conform to the rationale of 
Malus' Law; i.e., that it is to be the comparison of intensity 
measurements made in series, distinguish this result from the usual of 
this law. 

A semiclassical, sometimes denoted `naive,' model of the EPR emissions 
has been suggested.  It takes the detection probability to be 
proportional to the received field intensity: $({\cal 
E}\cos(\theta))^2$ where $\theta$ is the angle between the 
polarization direction of the signal and the axis of the polarizer 
used in the detector.  A coincidence detection is proportional to the 
product of a detection probabilities in each channel.   The resultant 
probability is proportional to: $\cos^2(\theta)\cos^2(\theta -\phi)$ 
where $\phi$ is the angle between the axes of the measurement 
polarizers if the coordinate system is aligned with one of them.  
(This is just Malus' Law generalized to a frame rotated by $\theta$.)
Finally, the total probability is obtained by averaging over many 
pairs of signals, each with its own randomly given polarization angle 
$\theta$, that is 

\beq 
{1\over{\pi}}\int_{0}^{\pi}[\cos(\theta)\cos(\theta - \phi)]^2 
d\theta = 1/4 +{1/8}cos(2\phi). 
\eeq 

To convert this intensity to a probability it must be converted to 
the ratio of coincidence rate divided by the total {\it pair} 
production rate.  For ideal detectors, the number of detections is 
linearly proportional to the field intensity so the number of pairs of 
detection events will be $1/2$ the value of the field intensity 
integrated over the detector apertures---in the above equations, this 
factor in the local units has been taken to be $1$, so that finally 
Eq. (16) must be divided by $1/2$.  

This expression does not violate a Bell Inequality.  It would be an 
ideal rebuttal to the conundrums evoked by Bell's [theory] were it to 
agree with experiment.  However, this result has a nonzero minimum 
whereas Eq. (15) does go to zero and this difference is observable; 
Eq. (16) does not conform to Nature. 
\cite{Clauser1} 

This simple observation would settle any dispute regarding the 
existence of a local realist alternative to QM were the above 
semiclassical model exhaustive.  In fact it is not---at least not if 
the projection postulate is abandoned.  A different result is obtained 
if to the above semiclasscial model the following two modifications 
are made: 

\begin{description} 
\item[a.\phantom{12345}] the source is assumed to be comprised of a 
cloud of independent point sources (e.g., atoms) which emit spherical 
radiation into $4\pi$ steradians (in their own rest frames) with 
random phase offsets, $\gamma$ determined by; {\it inter alia}, random 
locations in the cloud, random motion and so on; and, 

\item[b.\phantom{12345}] it is assumed that although a two stage cascade 
emission is facilitated by an intermediate step, that the radiation 
from each stage alone is unpolarized even if each stage seems to have 
a distinct frequency.  This means that the radiation from each source 
(atom) received in each detector will comprise two independent random 
contributions, one each from each polarization state.  
\end{description} 

The phase difference between each polarization state is a random 
function of time.  The signal received at the detector aligned with 
the coordinate system is proportional to two terms, one for each 
polarization state with its own phase offset, $\gamma_{x,y}$: 
 
\beq 
A(\theta, \gamma_x, \gamma_y) =\cos(\theta)\cos(\gamma_x) + 
\sin(\theta)\cos(\gamma_y),
\eeq 

and likewise, the signal at the detector oriented at angle $\phi$ to 
the other detector is proportional to: 

\beq 
\lefteqn{B(\theta, \phi, \gamma_x, \gamma_y) = } \nonumber \\ 
 & & \cos(\theta + \phi)\cos(\gamma_x) + \sin(\theta + 
\phi)\cos(\gamma_y). 
\eeq 

The coincidence count registered by ideal detectors will be then 
proportional to the average over all incidence angles, $\theta$, of 
the squared product of these two factors averaged over the phase 
offsets, $\gamma_{x,y}$: 

\beq
{1\over{\pi}}\!\!\int_0^{\pi}\!\![{1\over{\pi^2}}\!\! 
\int_0^{\pi}\!\!\!\!\int_0^{\pi}\!\!\!\!\!\! A(\theta, \gamma_x, 
\gamma_y) B(\theta, \phi, \gamma_x, \gamma_y)d\gamma_x d\gamma_y ]^2 
d\theta. 
\eeq
This integral, divided again by the ideal {\it pair} production rate,  
evaluates to: 
\beq 
{1\over{2}}\sin^2(\phi). 
\eeq 

Thus, in conclusion, this purely realist-local formulation with 
erstwhile hidden variables $\theta, \gamma_{x}$ and $\gamma_{y}$ 
yields a result which is identical to that from QM.  What allows this 
calculation to go through is that it violates the projection 
hypothesis by admitting not only more than two values of the field 
intensity at the detectors, but also values greater than $1$, namely 
the maximum value of $sin(\theta) + cos(\theta)$, i.e., $\sqrt{2}$ as 
the sum of two orthogonally polarized signals.  The factor  
$\sqrt{2}$ is recognized as that by which QM violates Bell 
inequalities. 

Abandoning a strict application of the projection hypothesis is not 
new.  Bell considered new definitions of the functions $A({\bf 
a},\lambda)$ and $B({\bf b},\lambda)$ which could take on the value 
zero in addition to $\pm 1$ to take into account failed detections of 
generated `particles.' \cite{Bell4} (Zero is not an eigenvalue of any 
Pauli spin operator.)  Then for these new functions a new inequality 
restricting their averages,  $\overline{A}({\bf a},\lambda)$ and 
$\overline{B}({\bf b},\lambda)$, was derived and used to show that a 
certain class of stochastic local hidden variables are incompatible 
with QM---given the hypothesis, the essential element of which here 
is being less than the norm of the eigenvalues of a spin operator: 
$1$.

Likewise, Clauser and Horne reformulated the derivation of Bell 
inequalities with the goal of obtaining versions that can be compared 
with experiments.  The basic problem they faced was that in order to 
obtain observable probabilities using the original Bell formulation, 
it is necessary to determine accurately the total production rate of 
pairs. However, if it is taken that some `particles' go undetected, 
then in principle measuring this rate accurately is impossible.  
Clauser and Horne evaded this problem by using only ratios of 
detection rates which are equal to ratios of detection probabilities 
as the normalizing factor, the total pair generation rate, 
fortuitously divides out.  

In the derivation of these altered inequalities, however, the 
projection postulate once again tacitly serves as motivation.  It 
enters in the assumption that a count-event is caused by an entity 
for which a dichotomic function specifies its state.  That is, it is 
assumed that, for example, a count is triggered by a photon with either 
pure $x$ or $y$ polarization, and not by some combination for which 
the norm could be greater than $1$.  The new local realist model 
described above violates exactly this assumption by assuming that the 
signal triggering a detection event (ejection of a photoelectric 
electron) has a contribution from each polarization state. 

Of course, the above new local realist model need not be evaluated 
exclusively under the umbrella of conceptual categories set by 
the projection postulate.  It stands alone as a counterexample to the 
claim that no local realist model can duplicate QM. 

\section{LOCALITY AND PROJECTION POSTULATE}

There are seemingly deep philosophical considerations regarding 
locality that, almost uniquely for such topics, are largely 
instinctively understood by any thoughtful person.  However 
interesting such matters may be, they are not within the purview of 
Physics which has is own requirements for locality.  The latter are 
expressed as the demand that equations of motion be mathematically 
well posed.  

From this requirement two degrees of `locality' can be discerned: what 
can be called `light induced' locality that requires all causes to be 
within the past light cone of their effects, and `general locality' 
which requires that such causes be within the past light-like cone 
where the speed of `interaction' is any thing below the limit  
$c\rightarrow\infty$.  Violating general locality, this author 
maintains\cite{EM2body}, results in poorly posed, unintegrable 
equations of motion and for this reason alone, strictly from within 
the discipline of physics, must be rejected.  In the zone between 
light and general locality, interactions faster than the speed of 
light can not be ruled out {\it a priori}.  Of course, until a new 
force is found for which the interaction speed exceeds that of light, 
or perhaps the discovery that the speed of gravity exceeds the speed 
of light,  all such considerations are exploratory. 

Thus, indications that QM reveals some kind of nonlocal interaction, 
need not be taken by itself as fundamentally unacceptable if what is 
meant does not violate general locality.  But, ``extraordinary claims 
require extraordinary evidence.''  Here, the  `extraordinary' cost of 
admitting the so far otherwise totally unsupported hypothetical 
existence of a new and superluminal force is much higher than 
abandoning the at best formalistic projection postulate. 

One might ask, exactly what purpose the projection postulate serves.  
For us today it is difficult to follow in detail the thought sequences 
of the originators, but it seems possible that they were focused on 
fitting distinct spectral lines into operators on Hilbert Spaces. 
\cite{Hughes} The effective assumption appears to have been that 
multiple lines seen when analyzing the emissions from a particular 
source arise separately from distinct atoms. This imagery has become 
enshrined in the notion that each atomic transition results in a 
distinct photon.  

However, the appearance of distinct spectral lines results whenever an 
emission results from a constrained system for which the relevant 
equations have compact support.  A guitar string, for example, when 
struck will generally not vibrate at a pure tone, be it the 
fundamental or any harmonic, but rather at some combination.  Spectral 
analysis, nevertheless, will reveal distinct lines---but one is not 
entitled to conclude that the signal is emitted by an ensemble of 
guitars, each sending a pure tone.  For all the same reasons, atomic 
spectra can not be attributed to the sum of pure signals from 
individual transitions in an atom so that an unrestricted application 
of the projection postulate is unwarranted, rendering the postulate 
itself questionable if not downright inconsistent.  Of course, 
reconciling new, alternate imagery with 70 years of precedence will 
take thought and time. 

\section{CONCLUSIONS}

When fully understood, what Bell's Theorems do do is prove just what 
their hypothesis allows, namely that a reformulation of QM with hidden 
variables will be unable to parameterize particle spin with dichotomic 
functions in directions different from the magnetic field.  At the 
heart of this issue is the tacit assumption that particles have 
dichotomic spin in arbitrary directions intrinsically, regardless of 
external magnetic fields, which, when introduced by observers, just 
`measure' the preexisting spin in the direction of the field. If, on 
the other hand, spin is seen as gyration actually engendered by the 
magnetic field, this whole issue and its attendant confusion do not 
arise. 

What these theorems do not do, however, is generally preclude 
hidden variable formulations.  Likewise, if the projection postulate 
is sacrificed in whole or part, they do not invest any part of Physics 
with nonlocal relationships of any kind, including for correlations. 

\end{document}